\documentclass[runningheads]{llncs}
\usepackage[T1]{fontenc}
\usepackage{graphicx}
\usepackage{amsmath} 
\usepackage{multirow}
\usepackage{booktabs} 
\usepackage{amssymb}
\usepackage[misc]{ifsym}
\begin{document}
\title{Continuous Dynamic Modeling via Neural ODEs for Popularity Trajectory Prediction}
%
%
\author{Songbo Yang \and
Ziwei Zhao \and
Zihang Chen \and
 Haotian Zhang \and
 Tong Xu \and
 Mengxiao Zhu \textsuperscript{(\Letter)}}
\institute{University of Science and Technology of China,Hefei, China\\ 
\email{   \{songboyang,zzw22222,czh1999,sosweetzhang\}@mail.ustc.edu.cn}\\
\email{\{ tongxu,mxzhu\}@ustc.edu.cn}
}
\maketitle              
\begin{abstract}
Popularity prediction for information cascades is a fundamental and challenging task in social network data analysis. While most existing methods consider this as a discrete problem, popularity actually evolves continuously, exhibiting rich dynamic properties. In this paper, we argue that popularity trajectory prediction is more practical, as it aims to forecast the entire continuous trajectory of how popularity unfolds over arbitrary future time. However, traditional methods for popularity trajectory prediction primarily rely on specific diffusion mechanism assumptions, which may not align well with real-world dynamics and compromise their performance. To address these limitations, we propose NODEPT, a novel approach based on neural ordinary differential equations (ODEs) for popularity trajectory prediction. We first employ an encoder to initialize the latent state representations of information cascades, consisting of two representation learning modules that capture the co-evolution structural characteristics and temporal patterns of cascades from different perspectives. More importantly, we then introduce an ODE-based generative module that learns the dynamics of the diffusion system in the latent space. Finally, a decoder transforms the latent state into the prediction of the future popularity trajectory. Our experimental results on three real-world datasets demonstrate the superiority and rationality of the proposed NODEPT method.

\end{abstract}
\section{Introduction}
With the growing popularity of social media, more users are sharing and interacting online, fundamentally changing how people entertain and communicate. Consequently, accurately predicting the popularity of information cascades on social media is essential for various applications, including viral marketing \cite{leskovecDynamicsViralMarketing2007}, assessing scientific impact \cite{wangQuantifyingLongTermScientific2013a}, and item recommendation \cite{wuDualSequentialPrediction2019}.

Popularity prediction aims to forecast the future number of users who will react to information. Early research efforts in this domain focused on proposing hypotheses about information diffusion mechanisms or utilizing manually engineered features to predict cascade popularity \cite{wangQuantifyingLongTermScientific2013a,maPredictingPopularityNewly2013}. As deep learning techniques advanced, a series of neural network-based methods were proposed. Many studies began employing recurrent neural network (RNN) based models to capture the structural properties and temporal dynamics of cascades  \cite{caoDeepHawkesBridgingGap2017,islamDeepDiffusePredictingWho2018,liaoPopularityPredictionOnline2019}. Nevertheless, these RNN-based approaches cannot fully leverage the structural information within cascades. To address this limitation, graph representation learning methods, such as graph neural network (GNN) based models, were introduced to further improve prediction performance by comprehensively learning cascade structures \cite{xuCasFlowExploringHierarchical2023,chengCasODEModelingIrregular2023,caoPopularityPredictionSocial2020}. Moreover, dynamic graph learning algorithms have been adopted to track the evolution of user preferences over time \cite{jiCommunitybasedDynamicGraph2023,luContinuoustimeGraphLearning2023}. 

\begin{figure}[ht]
    \centering
    \includegraphics[width=0.5\textwidth]{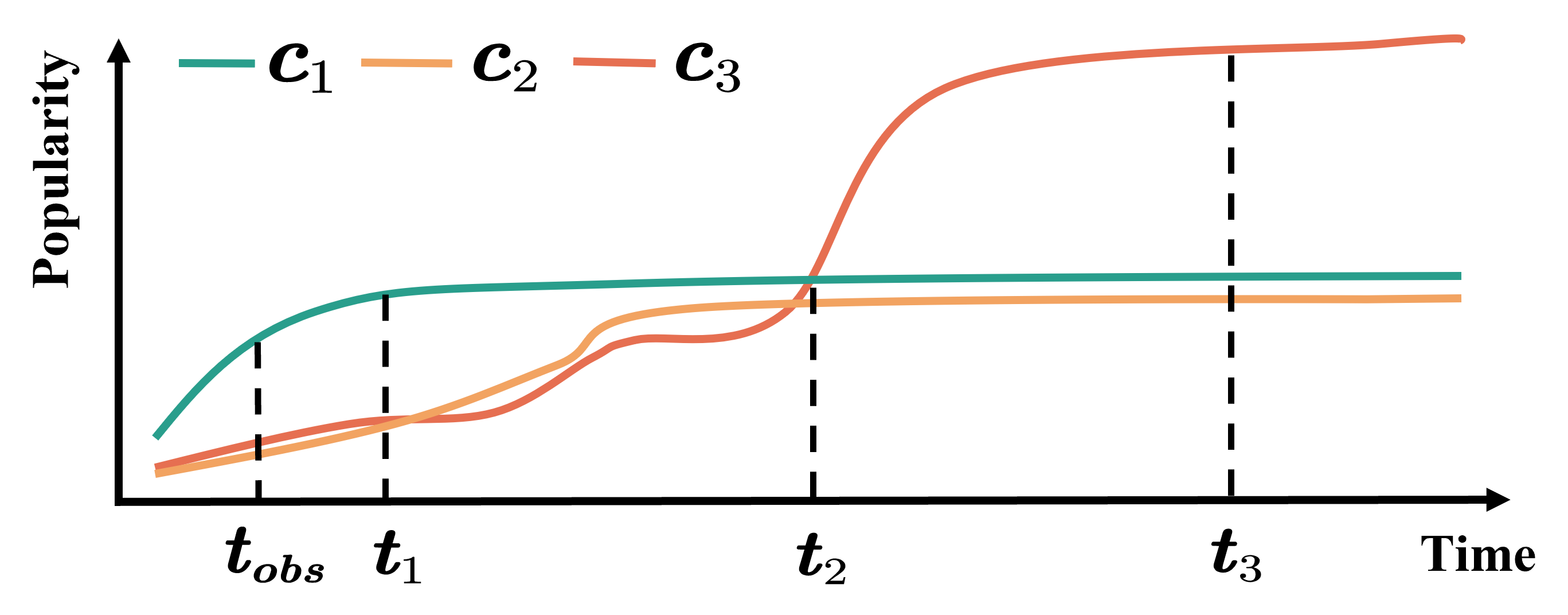}
    \caption{ Popularity trajectories of different information cascades.}
    \label{Illustration of trajectory prediction}
\end{figure}

Despite the promising performance of these models, they still face key limitations. Notably, the majority of current methods typically model popularity prediction as a single point prediction problem, forecasting incremental popularity after a fixed period $\Delta t$ \cite{luContinuoustimeGraphLearning2023,xuCasFlowExploringHierarchical2023,jiCommunitybasedDynamicGraph2023}. To predict popularity at different timestamps, these models require retraining \cite{zhangAnytimeInformationCascade2022}, which increases computational costs. Some models approach popularity prediction as a time series forecasting problem \cite{wangAFRFAngleFeature2023}, but they still use discrete methods, predicting at specific points rather than continuously.  In reality, the popularity of information cascades evolves continuously, exhibiting valuable dynamic properties such as change rates and growth patterns. Discrete methods inadequately capture these dynamics.
We argue that popularity trajectory prediction provides a more practical and informative approach, as it forecasts the entire trajectory of how popularity unfolds over arbitrary future time points. This approach offers insights into both instantaneous popularity and underlying dynamics. 
For example, Figure \ref{Illustration of trajectory prediction} shows that at time $t_1$, although the popularity of cascade $c_1$ is significantly higher than that of $c_2$ and $c_3$, its change rate is lower. By the time $t_2$, while the popularity levels of $c_1$, $c_2$, and $c_3$ are similar, the change rate of $c_3$ is significantly greater. 
Understanding change rates helps in crafting better-targeted interventions, such as boosting positive content or cutting off misinformation cascades during critical periods.

Traditional popularity trajectory prediction models often depend on diffusion mechanism assumptions, such as the self-exciting Hawkes processes \cite{zhangAnytimeInformationCascade2022,zhaoSEISMICSelfExcitingPoint2015}. However, these predefined assumptions limit their ability to capture the complex patterns of popularity growth fully and may not align well with real-world dynamics, which can compromise their performance. Therefore, developing a data-driven neural network approach is crucial. Such methods learn complex patterns directly from the data and capture intricate characteristics that traditional methods might overlook, leading to more accurate predictions.


To address the above limitation, we propose a novel neural ordinary differential equations (ODEs)-based method for popularity trajectory prediction, named NODEPT. Generally, NODEPT models the dynamics of the underlying diffusion system via neural ODEs.
 Firstly, we construct a dynamic diffusion graph and a dynamic graph representation learning module is proposed to extract the co-evolution structural characteristics of cascades. Secondly, a temporal sequence representation learning module sequentially aggregates user dynamic representations to learn the temporal patterns within each cascade. Thirdly, we fuse the representations obtained from these representation learning modules into a unified cascade representation, inferring the initial latent state for each cascade. More importantly, we then propose an ODE-based generative module to capture the evolution of the diffusion system. The ODE function incorporates the latent state of cascades and their interaction representation that is obtained by adaptively aggregating relevant cascade information from a memory module. Finally, a decoder transforms the latent continuous trajectory produced by the ODE generative module into the future popularity trajectory for each cascade.

Our main contributions can be summarized as follows:
\begin{itemize}
\item Different from traditional methods, we explore a continuous data-driven network approach without predefined diffusion mechanisms, which captures intricate characteristics that might missed by traditional methods, offering more accurate predictions for continuous popularity trajectory.

\item To fully leverage neural networks for extracting latent information from data, we propose NODEPT, a novel framework based on neural ODEs for continuous popularity trajectory prediction. NODEPT integrates an ODE-based generative module to track the evolution of the underlying diffusion system. Our approach effectively captures both the self-evolution of individual cascades and their complex interactions with other cascades.

\item Extensive experiments on multiple real-world datasets demonstrate the effectiveness of NODEPT in popularity trajectory prediction, validating its superior performance compared to traditional methods.
\end{itemize}


\section{Related work}
\subsection{Popularity  (Trajectory) Prediction}

The popularity prediction task for information cascades aims to forecast how many users will react (i.e., comment, like, share, or retweet) to a given information. This problem has been extensively studied by researchers. 
Some early efforts focused on feature-based methods that utilize manually engineered features to predict cascade popularity \cite{chengCanCascadesBe2014,maPredictingPopularityNewly2013,maWillThisHashtag2012}.

With the emergence of deep learning,  some methods extracted diffusion sequences from cascades and employed sequence models such as recurrent neural networks (RNNs) to capture structural and temporal patterns \cite{caoDeepHawkesBridgingGap2017,islamDeepDiffusePredictingWho2018,liaoPopularityPredictionOnline2019}. For instance, DFTC \cite{liaoPopularityPredictionOnline2019} utilized CNNs and LSTMs to derive node embeddings from popularity count sequences. Despite achieving good performance, these sequence-based methods still failed to capture the underlying cascade graph structures fully.
To address this limitation, some recent methods leveraged graph representation learning techniques to comprehensively model both structural and temporal information \cite{chenInformationDiffusionPrediction2019,xuCasFlowExploringHierarchical2023,luContinuoustimeGraphLearning2023,chengInformationCascadePopularity2024}. For instance, CTCP \cite{luContinuoustimeGraphLearning2023} used dynamic graph learning to capture user interest evolution and LSTMs to model temporal patterns.   However, these methods remained discrete approaches that predicted at specific points rather than continuously, which limited their ability to provide flexible and informative predictions.

Popularity trajectory prediction aims to forecast the entire trajectory of how popularity unfolds over arbitrary future time points. Traditional methods primarily relied on specific diffusion mechanism assumptions and characterizing cascades using generative approaches including the difference-based model \cite{xuEpidemicInformationDissemination2015,staiTemporalDynamicsInformation2018}, various stochastic point processes \cite{zhangAnytimeInformationCascade2022,zhaoSEISMICSelfExcitingPoint2015}. For example, CASPER \cite{zhangAnytimeInformationCascade2022} assumed that the information diffusion process follows marked Hawkes point processes and presents closed-form expressions for the mean and variance of future event counts. However, these specific diffusion mechanism assumptions may not fully capture the underlying patterns governing the growth of information cascade popularity.  In this paper, we propose a novel neural ODE-based method to model the continuous dynamics of the diffusion system that automatically captures complex patterns from data. 
 

\subsection{Neural Ordinary Differential Equation}

Neural ordinary differential equations (ODEs) \cite{Chen2018NeuralOrdinaryDifferential} define a family of continuous-time models where a hidden state $h(t)$ evolves according to an ODE initial value problem:

\begin{equation}
\frac{dh(t)}{dt}=f_\theta(h(t),t)\quad\mathrm{where} \, h(t_0)=h_0,
\end{equation}
in which the function $f_\theta$ specifies the dynamics of the hidden state, using a neural network with parameters $\theta$. Given the initial states $h(t_0)$, and the ODE function $f$, any black box numerical ODE solver such as Runge-Kutta \cite{schoberProbabilisticModelNumerical2019} can solve the ODE initial-value problem (IVP). The solution  can then be evaluated at any desired time $t$ as:
\begin{equation}
    h(T)=h(t_0)+\int_{t=t_0}^Tf_\theta(h(t),t)dt.
\end{equation}

Due to their flexibility and capabilities, neural ODEs have been widely adopted across various research fields such as graph neural networks \cite{xhonneuxContinuousGraphNeural2020} and popularity prediction  \cite{chengCasODEModelingIrregular2023,chengInformationCascadePopularity2024}. 
In the domain of popularity prediction,  
CasODE \cite{chengCasODEModelingIrregular2023} and CasDO \cite{chengInformationCascadePopularity2024} generalized discrete RNN state transitions to continuous-time dynamics for modeling irregularly sampled cascade events. However, they still focused on single-point popularity prediction rather than the full continuous popularity trajectory prediction.


\section{Preliminaries}
\textbf{Cascade.} Given a set of users $\mathcal{U}$ , a cascade $c$ records the diffusion process of a information  among the users $\mathcal{U}$. Specifically, we use a chronological sequence $g^c(t)=\{(u_j^c,v_j^c,t_j^c)\}_{j=1,...,|g^c(t)|}$ to represent the growth process of cascade $c$ until time $t$, where $(u_j^c,v_j^c,t_j^c)$ indicates that  $u_j^c$ shared information to $v_j^c$ at time $t_j^c$ (or equivalently, that user $v_j^c$ participates in  cascade $c$ due to user $u_j^c$ at $t_j$).

\textbf{Diffusion Graph.} A diffusion graph is defined as a chronological sequence of diffusion behavior $\mathcal{G}^t=\{(u_j,v_j,c_j,t_j)|t_j<t\}$ to denote the diffusion process of all cascades until $t$, where $(u_{j},v_{j},c_{j},t_{j})$ represents user $v_j$participates in cascade $c_j$ due to user $u_j$ at $t_j$.

\textbf{Diffusion System.} A diffusion system is a dynamic system that describes the dynamic process by which information spreads through a network of users over time. At any given time $t$, the overall state of the diffusion system can be represented by the set of all latent state $\{\boldsymbol{z}_c^t|c \in \mathcal{C}\}$, where $\boldsymbol{z}_c^t$  represents the continuous latent state variable encapsulating the diffusion status of cascade c at time t and $\mathcal{C}$ denotes the set of all cascades.

\textbf{Popularity Trajectory Prediction.} Given a cascade $c$ begins at $t_0^{c}$, after observing it up to time $t_{obs}^c$
,  our goal is to learn the underlying dynamics which is built upon the latent representations for cascade $\boldsymbol{z}_c^t\in \mathbb{R}^d$, and to utilize it to predict the trajectory of incremental popularity $\hat{P_c^t}=|g^{c}(t)|-|g^c(t_{obs}^c)|$  of $c$   in the future, where $t>t_{obs}$. 
In particular, we learn a function $f: \mathcal{G}^{t_{obs}^c} \to \hat{P_c^t}$. Here, $\hat{P}_c^t$ can be understood as a function of time $t$, where its output represents the incremental popularity of cascade $c$ at time $t$.






\section{Method}
\begin{figure*}
    \centering
    \includegraphics[width=\textwidth]{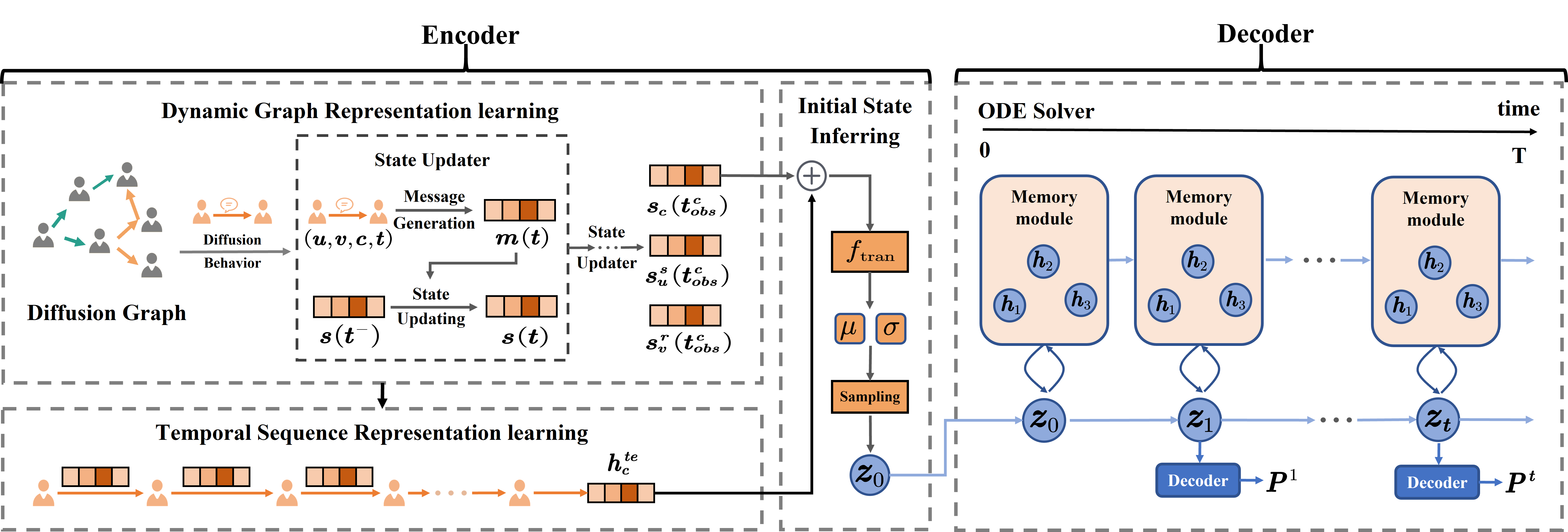}
    \caption{Framework of our proposed NODEPT. }
    \label{fig:model}
\end{figure*}

The overall NODEPT framework is illustrated in Figure \ref{fig:model}.   Following the variational autoencoder (VAE) paradigm, NODEPT consists of three jointly-trained components:
(1) An initial states encoder that models co-evolution structural characteristics and temporal patterns of cascades for initializing latent state representations (2) A generative module characterized by an ODE function for modeling the evolution of the underlying diffusion system in the latent space. (3) A decoder that generates the predicted trajectory based on the decoding likelihood determined by the latent states $p\left(\hat{P_c^t}\mid z_c^t\right)$.

\subsection{Initial State Encoder}

Given the diffusion graph $\mathcal{G}^{t^c_{obs}}$ up to the observation time $t^c_{obs}$ for cascade $c$, the initial state encoder computes a posterior distribution $q_\phi(\boldsymbol{z}_c^0|\mathcal{G}^{t^c_{obs}})$ of the latent initial state $\boldsymbol{z}_c^0$, from which a sample is drawn.
Firstly, we employ dynamic graph representation learning method to extract the co-evolution structural characteristics of cascades. Secondly, we learn the temporal patterns of a cascade by leveraging the diffusion sequence to aggregate the user dynamic representations over time. Subsequently, we combine the dynamic representation and temporal representation of cascade $c$ into a unified cascade representation $\boldsymbol{h}_c$, thereby inferring the initial state $\boldsymbol{z}_c^0$ for cascade $c$. 
\subsubsection{Dynamic Graph  Representation Learning}
Inspired by the continuous-time dynamic graph learning method \cite{rossiTemporalGraphNetworks2020}, we dynamically learn the diffusion behaviors within the diffusion graph  $\mathcal{G}^{t^c_{obs}}$ as they occur.
To model the evolutionary patterns of users and cascades, we maintain dynamic states for both users and cascades. For each user $u$, we maintain two types of states $\boldsymbol{s}_{u}^{s}(t)$ and $\boldsymbol{s}_{u}^{r}(t)$ to adequately capture their roles as information senders and receivers throughout the diffusion process. Additionally, for each cascade c, we maintain a dynamic cascade state $\boldsymbol{s}_c(t)$.

To effectively integrate information from the previous states and the current diffusion behavior, we first encode the diffusion behavior $(u, v, c, t)$ into message representations $\boldsymbol{m}_u(t)$, $\boldsymbol{m}_v(t)$, and $\boldsymbol{m}_c(t)$ for the users and cascade respectively.
Taking the cascade state $\boldsymbol{s}_c(t)$ as an example, let $\boldsymbol{s}_c(t^-)$ denote the cascade state just before time $t$. We generate the message representation for cascade $c$ using the following mechanism:

\begin{equation}
    \boldsymbol{m}_c(t)=\sigma(\boldsymbol{W}^m[\boldsymbol{s}_u^s(t^-)||\boldsymbol{s}_v^r(t^-)||\boldsymbol{s}_c(t^-)||\boldsymbol{f}_c^t]+\boldsymbol{b}^m),
    \label{massage}
\end{equation}
where || is the concatenation operation and $\boldsymbol{W}^m \in \mathbb{R}^{d \times 3d}, \boldsymbol{b}^m \in \mathbb{R}^d$ are trainable parameters, $\boldsymbol{f}_c^t$ denotes the temporal feature generated from timestamp $t$ . Inspired by \cite{xuInductiveRepresentationLearning2020},  $\boldsymbol{f}_c^t$ takes the following form,
\begin{equation}
    \boldsymbol{f}_c^t=[\cos w_1^r\Delta t_c,\cos w_2^r\Delta t_c,...,\cos w_n^r\Delta t_c],
\end{equation}
where $\Delta t_{c}$ is the time interval since the last updating of cascade $c$ (i.e., $\Delta t_{c}=t-t_{c}^{-}$ and $t_c^-$ is the last time when $c$ was updated), $\boldsymbol{w}=[w_{1},...,w_{n}] \in \mathbb{R}^d$ are trainable parameters.
After generating message representations, we fuse it with the previous cascade state $\boldsymbol{s}_c(t^-)$ using a Gated Recurrent Unit (GRU) to capture the diffusion history and evolutionary patterns as follows.
\begin{equation}
    \boldsymbol{s}_{c}{(t)}=GRU\left( \boldsymbol{s}_{c}(t^-),\boldsymbol{m}_c(t)\right).
    \label{dynamic state update}
\end{equation}
The strengths of GRU in capturing long-range temporal dependencies make it well-suited for this continuous state evolution modeling.

The user sender $\boldsymbol{s}_u^s(t)$ and receiver $\boldsymbol{s}_u^r(t)$ states are updated similarly, fusing their respective messages $\boldsymbol{m}_u(t)$ and $\boldsymbol{m}_v(t)$ with previous states using GRU.

\subsubsection{Temporal Sequence  Representation Learning}
In this module, we learn the temporal patterns of a cascade by leveraging the diffusion sequence to aggregate the user dynamic states over time.  Given a cascade $c$,  we first organize cascade c as a diffusion sequence of user-time pairs $\{(u_1^c, t_1^c), (u_2^c, t_2^c), ..., (u_n^c, t_n^c)\}$, where $u_j^c$ is the user and $t_j^c$ is the time when user $u_j^c$ shared the information and participate in the cascade $c$.  To enhance the model's ability to perceive the position information of user shares,
 for each user $u_i^c$ in the sequence, we compute their temporal user embedding $\boldsymbol{h}_i^c$ by first adding  position embedding to the user' dynamic state representation:
\begin{equation}
    \boldsymbol{h}_i^c=\boldsymbol{s}_u^i+\boldsymbol{e}^p_i,
    \label{h_ui}
\end{equation}
where $\boldsymbol{s}_u^i$ is the dynamic state $\boldsymbol{s}_u^s(t_{obs}^c)$ of user $u_i^c$ and  $\boldsymbol{e}^p_i$ is position embedding.

We then feed the sequence of temporal user embeddings $\{\boldsymbol{h}_1^c $, $\boldsymbol{h}_2^c, ..., \boldsymbol{h}_n^c\}$ into a Long Short-Term Memory (LSTM) network to produce the temporal representation $\boldsymbol{h}_c^{te}$ of cascade $c$.
\begin{equation}
    \boldsymbol{h}_c^{te}=LSTM([\boldsymbol{h}_1^c,\boldsymbol{h}_2^c,...,\boldsymbol{h}_n^c]).
    \label{tem representation}
\end{equation}

\subsubsection{Initial State Inferring}


Whenever the observed time $t_{obs}^c$ of cascade $c$ is reached, we first compute the dynamic representation of the cascade. To capture the  characteristic of  the publishing time $t_0^c$ of the cascade, we then divide the overall observation period $[0, T_{obs}]$ into $N$ equal time slots. For each time interval $[k\frac{T_{obs}}{N}, (k+1)\frac{T_{obs}}{N})$, we have a learnable embedding $\boldsymbol{e}^{dy}_k$ to distinguish cascade published in different time intervals. Next, we obtain the dynamic representation of the cascade as follows.
\begin{equation}
    \boldsymbol{h}_c^{dy} = \boldsymbol{s}_c(t_{obs}^c)+\boldsymbol{e}^{dy}_{[t_0^c]},
    \label{h_c_dy}
\end{equation}
where $\boldsymbol{s}_c(t_{obs}^c)$ is the dynamic state of cascade c at the observation time $t_{obs}^c$, and $\boldsymbol{e}^{dy}_{[t_0^c]}$ is time slot embedding  that encodes the publishing time $t_0^c$ of the cascade.

The dynamic representation $\boldsymbol{h}_c^{dy}$ and temporal representation $\boldsymbol{h}_c^{te}$ explore the rich cascade information from different perspectives.
To generate the overall cascade representation $\boldsymbol{h}_c$ for initializing the latent state, we concatenate these two representations and then feed them through a multi-layer perceptron (MLP):

\begin{equation}
    \boldsymbol{h}_c=\sigma \left( \boldsymbol{W^i} \left[ \boldsymbol{h}_c^{dy}||\boldsymbol{h}_c^{te} \right] \right),
\end{equation}
where $\boldsymbol{W^i} \in \mathbb{R}^{d \times 2d}$ are learnable parameters and $\sigma(\cdot)$ is a non-linear activation function to provide non-linearity. 

Afterward, we utilize $\boldsymbol{h}_c$ to infer the approximate posterior distribution $q_\phi\left(\boldsymbol{z}_c^0\mid  \mathcal{G}^{t^c_{obs}} \right)$ from which we sample the initial latent state $\boldsymbol{z}_c^0$ for the ODE generative model.

\begin{equation}
\begin{gathered}
    \boldsymbol{\mu}_{z_{c}^{0}},\boldsymbol{\sigma}_{z_{c}^{0}}=f_{\mathrm{trans}}\left(\boldsymbol{h}_{c}\right),\\
    q_\phi\left(\boldsymbol{z}_c^0\mid \mathcal{G}^{t^c_{obs}}  \right)=\mathcal{N}\left(\boldsymbol{\mu}_{z_c^0},\boldsymbol{\sigma}_{z_c^0}\right),
\end{gathered}
\end{equation}
where $\mathcal{N}$ denotes a normal distribution, $f_{\mathrm{trans}}$ is a simple MLP whose output vector is equally split into two halves to represent the mean and variance respectively.
We then employ the reparametrization trick to sample from the posterior distribution $\boldsymbol{z}_c^0$.
\begin{equation}
    \boldsymbol{z}_c^0\sim p(\boldsymbol{z}_c^0)\approx q_\phi\left(\boldsymbol{z}_c^0|\mathcal{G}^{t^c_{obs}}\right).
    \label{initial state}
\end{equation}
Here $p(z_0)$ is a prior distribution used as a regularization. We minimize the KL divergence between $q_\phi\left(\boldsymbol{z}_c^0 | \mathcal{G}^{t_{obs}^c}\right)$ and the prior distribution $p(\boldsymbol{z}_0)$ during training.

\subsection{ODE Generative Module and Decoder}
\subsubsection{ODE Generative Module.}

After establishing the initial state encoder, we now define the neural ODE function that drives the diffusion system forward. The latent state of each cascade is influenced by two crucial factors: 1)  the self-evolution dynamics of the cascade itself, and 2) the potential influence received from interactions with other cascades. Therefore, our ODE function consists of two parts: one for the cascade’s self-evolution and one for capturing interactions with other cascades.

In the real world, new information and diffusion behaviors continually emerge. Therefore, determining which cascades interact with cascade $c$ and
the intricate interactions between them within the diffusion system is a key challenge. Since the influence of information cascades decays over time, cascades observed much earlier will not interact with the current cascade.  Thus, we propose an external memory module to store representations of the most recent historical cascades up to $t^c_{obs}$.  
This module maintains a memory matrix $\mathbf{M}_c \in \mathbb{R}^{N_m \times d}$, where $N_m$ is the memory module size and $d$ is the representation dimension.

We consider the cascades in $\mathbf{M}_c$ as potentially interacting with the current cascade $c$ and dynamically update the memory module as new information emerges. Moreover, we employ an attention mechanism to model these interactions by adaptively aggregating information from $\mathbf{M}_c$.
In formulation, we have:
\begin{equation}
    \begin{gathered}
    \operatorname{Att}(\mathbf{Q},\mathbf{K},\mathbf{V})=\operatorname{softmax}\left(\mathbf{QK}^T\right)\mathbf{V},\\
    h_c^{t,in}= \operatorname{Att}\left(\boldsymbol{z}_c^t W^Q, \mathbf{M}_cW^K,\mathbf{M}_cW^V  \right),
\end{gathered}
\end{equation}
where $z_c^t$ is the latent state of cascades, $W^Q \in\mathbb{R}^{d \times 2d},W^K \in \mathbb{R}^{d \times 2d},W^V \in \mathbb{R}^{d \times d}$ are trainable parameters and $h_c^{t,in}$ is the interaction representation of cascade $c$ with other cascades.
This mechanism can be viewed as a function $f_a \colon (\boldsymbol{z}_c^t, \mathbf{M}_c) \to h_c^{t, in}   $ that maps the latent state $\boldsymbol{z}_c^t$ of cascade $c$ and the matrix $\mathbf{M}_c$  to their interaction representation.

Furthermore, we define the overall neural ODE function as:

\begin{equation}
    \begin{gathered}
        \boldsymbol{h}_c^{t,in} =f_a\left(\boldsymbol{z}_c^t,\mathbf{M}_c\right),\\
        \frac{d\boldsymbol{z}_c^t}{dt}=f_g\left(\boldsymbol{z}_c^t,\mathbf{M}_c\right)=\sigma\left( \boldsymbol{W^f}\left[\boldsymbol{h}_c^{t,in} \Vert \boldsymbol{z}_c^t\right]\right),
    \end{gathered}
    \label{ODE function}
\end{equation}
where $\boldsymbol{W^f} \in \mathbb{R}^{d \times d} \in $ is trainable parameters. Here, $\boldsymbol{z}_c^t$ models the self-evolution of $c$ itself, while $\boldsymbol{h}_c^{t, in}$ captures the intricate interaction signals between $c$ and other cascades.





\subsubsection{Decoder}

Given the  ODE function $f_g$ and the initial state $\boldsymbol{z}_c^0$ of cascade $c$, the popularity trajectory of  $c$ is determined, which can be solved via a black-box ODE solver. Finally, a decoder generates the predicted trajectory based on the decoding probability $p({P_c^{t}}|\boldsymbol{z}_c^{t})$ computed from the decoding function $f_{\mathrm{dec}}$ as shown below. 
\begin{equation}
    \hat{P_c^t} \sim p(P_c^t|\boldsymbol{z}_c^t)=f_{\mathrm{dec}}(\boldsymbol{z}_c^t).
    \label{decoder}
\end{equation}
We implement $f_{\mathrm{dec}}$ as a  two-layer MLP with nonlinear activation. It outputs the mean of the normal distribution $p({P_c^{t}}|\boldsymbol{z}_c^{t})$, which we treat as the predicted popularity value for cascade $c$ at time $t$.

\subsection{Prediction and Training}
\subsubsection{Prediction}
We now introduce the overall prediction procedure of NODEPT. For each cascade, we condition its and other cascades' diffusion behaviors that occurred before the observation time i.e. $ \mathcal{G}^{t_{obs}^c}$ in order to predict its popularity trajectory in the future interval  $[t_{obs}^c, t_{pre}^c]$. Given the diffusion graph $ \mathcal{G}^{t_{obs}^c}$, we firstly run the encoder to compute the posterior distribution $q_\phi\left(\boldsymbol{z}_c^0\mid \mathcal{G}^{t_{obs}^c}  \right)$ for cascade $c$. Once the observation time is reached, we sample the initial latent states $\boldsymbol{z}_c^0$  from the posterior distribution. We then run the generative model defined by the ODE function to compute the latent state for cascade $c$ in the future. Next, we run the decoder to compute the mean of each decoding distribution as $\mu_{c}^{t}=f_{\mathrm{dec}}(\boldsymbol{z}_{c}^{t})$  which is treated as the predicted popularity for $c$ at time $t$, i.e., $\hat{P_c^t}=\mu_{c}^{t}$. 

\subsubsection{Joint training}
 The standard VAE framework is trained to maximize the evidence lower bound (ELBO). Inspired by \cite{huangCoupledGraphODE2021}, we first derived the ELBO loss for NODEPT as shown in Equation \ref{ELBO}.
 The first term is the reconstruction loss for the predicted trajectories and the second term is the KL divergence regularizing the inferred posteriors towards a prior $p(\boldsymbol{z}_0^c)$.

\begin{equation}
    \begin{aligned}
        \mathcal{L}_{ELBO\left(\theta,\phi\right)}&=\frac1N\sum_{c}ELBO_c\left(\theta,\phi\right),\\
        &=\frac1N\sum_{c}\mathcal{RL}-\mathrm{KL}\left[q_{\phi}\left(\boldsymbol{z}_{c}^{0}|\mathcal{G}^{t_{obs}^c}\right) \Vert p\left(\boldsymbol{z}_{c}^{0}\right)\right],\\
        &=\frac1N\sum_{c}\mathbb{E}_{z_{c}^{0} \sim q_{\phi}\left(\boldsymbol{z}_{c}^{0}|\mathcal{G}^{t_{obs}^c}\right)}[\log p_{\theta}\left(P_c^{t_{obs}^c \colon t_{pre}^c}\right)]
        -\mathrm{KL}\left[q_{\phi}\left(\boldsymbol{z}_{c}^{0}|\mathcal{G}^{t_{obs}^c}\right) \Vert p\left(\boldsymbol{z}_{c}^{0}\right)\right].
    \end{aligned}
    \label{ELBO}
\end{equation}
The reconstruction loss is estimated as below where the constant $\sigma$ is the standard derivation of each prior distribution.
\begin{equation}
    \mathcal{RL}=-\sum_t\frac{\Vert P_c^t-\hat{P_c^t} \Vert ^2}{2\sigma^2_2}.
    \label{reconstruction loss}
\end{equation}

Subsequently, recognizing that the reconstruction loss constrains popularity values only at discrete time points, we propose that an effective popularity prediction trajectory should also exhibit increments that align with the true trajectory between these points. Specifically, these increments reflect the growth pattern of the trajectory. To capture this growth pattern, we introduce an increment loss, defined in Equation \ref{increment loss}, which measures the discrepancy between the increments of adjacent time points in both the true and predicted trajectories.

\begin{equation}
    \begin{gathered}
        \mathcal{PI}=\hat{P_c^{t+1}}-\hat{P_c^t},\\
        \mathcal{TI}=P_c^{t+1}-P_c^t,\\
        \mathcal{IL}=\frac1N\sum_c\sum_t|\mathcal{PI}- \mathcal{TI}|.
    \end{gathered}
    \label{increment loss}
\end{equation}
Finally, we combined the two losses mentioned above to obtain the overall model loss and jointly trained the encoder, ODE generative model, and decoder in an end-to-end manner.
\begin{equation}
    \mathcal{L}=\lambda_1\mathcal{IL}-\mathcal{L}_{ELBO\left(\theta,\phi\right)}.
    \label{loss}
\end{equation}

\section{Experiments}

In this section, we conduct experiments on three real-world datasets to evaluate the effectiveness of our method, which tries to answer the following research questions (RQs).
\begin{itemize}
    \item \textbf{RQ1.} How does NODEPT perform compared to baseline models?
    \item \textbf{RQ2.} How do different components of NODEPT affect its performance?
    \item \textbf{RQ3.} How do the different hyperparameter settings affect the performance?
    \item \textbf{RQ4.} How does NODEPT perform for popularity trajectory prediction and what insights can be reflected from the trajectory?
\end{itemize}

\subsection{Experimental Settings}

\subsubsection{Descriptions of Datasets}
\begin{table}[h!]
\centering
\caption{Statistics of datasets.}
\label{Statistics of datasets.}
\begin{tabular}{lrrr}
\hline
Datasets & \#Users & \#Cascades & \#Retweets \\
\hline
Twitter & 199,005 & 19,718 & 602,253 \\
\hline
Weibo & 918,852 & 39,076 & 1,572,287 \\
\hline
APS & 218,323 & 48,575 & 939,686 \\
\hline
\end{tabular}
\end{table}

The experiments are conducted on three real-world datasets, including Twitter \cite{wengViralityPredictionCommunity2013}, Weibo \cite{caoDeepHawkesBridgingGap2017}, and APS \footnote{https://journals.aps.org/datasets}  that have been commonly used in previous related works \cite{luContinuoustimeGraphLearning2023,xuCasFlowExploringHierarchical2023,chengInformationCascadePopularity2024} for evaluating cascade popularity prediction. Table \ref{Statistics of datasets.} shows the statistics of the datasets.

\subsubsection{Baselines}
To assess the effectiveness of our method in single-point popularity prediction task, we compare our model with a variety of baselines.  \textbf{Xgboost} \cite{chenXGBoostScalableTree2016} is a scalable end-to-end tree boosting system, which is widely used in machine learning field. \textbf{DeepHawkes} \cite{caoDeepHawkesBridgingGap2017} and  \textbf{CasCN} \cite{chenInformationDiffusionPrediction2019} utilize RNNs to capture structural and temporal patterns in cascades.  \textbf{CasFlow} \cite{xuCasFlowExploringHierarchical2023} and \textbf{CTCP} \cite{luContinuoustimeGraphLearning2023}) employ 
 RNNs to learn the temporal features of cascades and utilize graph representation learning algorithms to effectively capture the structural features of cascades.  \textbf{CasODE} \cite{chengCasODEModelingIrregular2023}, \textbf{CasDO} \cite{chengInformationCascadePopularity2024} focus on modeling the irregular cascade events using  Neural ODEs for prediction. Meanwhile, We compare NODEPT with  \textbf{SEISMIC} \cite{zhaoSEISMICSelfExcitingPoint2015} and  \textbf{CASPER} \cite{zhangAnytimeInformationCascade2022} for the popularity trajectory prediction task. \textbf{SEISMIC} \cite{zhaoSEISMICSelfExcitingPoint2015} and \textbf{CASPER} \cite{zhangAnytimeInformationCascade2022} assume that the information diffusion process follows marked Hawkes point processes and \textbf{CASPER} is the state-of-the-art method for popularity trajectory prediction.


\subsubsection{Implementation Details}
We randomly divide the cascades into three sets: 70\% for training, 15\% for validation, and 15\% for testing purposes. We set the observation length of a cascade to 2 days, 2 hours, and 2 years on Twitter, Weibo, and APS.  Correspondingly, the prediction length of a cascade is set to 13 days, 13 hours, and 13 years for Twitter, Weibo, and APS datasets, respectively. 
We set the dimension of dynamic states of users and cascades as well as cascades representation to 64.  The time slot number $N$ in Equation \ref{h_c_dy}  is set to 50.  The memory module size is set to 16.
We set the standard derivation of prior distribution $\sigma$ in Equation \ref{reconstruction loss} as  0.1. The $\lambda_1$ in Equation \ref{loss} is set to 50.
For training, we employ the Adam optimizer with an initial learning rate of 0.0001. Batch sizes are set at 64 for Twitter and APS datasets and 128 for Weibo. We evaluate model performance using Mean Squared Logarithmic Error (MSLE) and Mean Absolute Percentage Error (MAPE). Our code can be found at 
\url{ https://anonymous.4open.science/r/ODEPT-0433}.

\begin{table}[htbp]
\centering
\caption{Model Performance of single-point prediction On Twitter, Weibo, APS datasets in terms of MSLE, MAPE}
\resizebox{\textwidth}{!}{
\begin{tabular}{lccccccccccccc}
\toprule
\multirow{3}{*}{Model} & \multicolumn{4}{c}{Twitter} & \multicolumn{4}{c}{Weibo} & \multicolumn{4}{c}{APS} \\
\cmidrule(lr){2-5} \cmidrule(lr){6-9} \cmidrule(lr){10-13} 
& \multicolumn{2}{c}{5 Days} & \multicolumn{2}{c}{15 Days} & \multicolumn{2}{c}{5 Hours} & \multicolumn{2}{c}{15 Hours} & \multicolumn{2}{c}{5 Years} & \multicolumn{2}{c}{15 Years} \\
\cmidrule(lr){2-3} \cmidrule(lr){4-5} \cmidrule(lr){6-7} \cmidrule(lr){8-9} \cmidrule(lr){10-11} \cmidrule(lr){12-13} 
& MSLE & MAPE & MSLE & MAPE & MSLE & MAPE & MSLE & MAPE & MSLE & MAPE & MSLE & MAPE \\
\midrule
Xgboost & 8.3209& 0.7719 & 10.1375 & 0.8478 & 3.1755  & 0.3261 & 4.0548 & 0.4139 & 2.2523 & 0.3314  & 4.2351 & 0.5863 \\
CasODE & 2.4529 & 0.2637  & 4.1891 &  0.3484 & 1.7659 & 0.2434 & 2.1098 & 0.2349 & 1.0852 & 0.2267 & 1.7357 & 0.2934 \\
DeepHawkes & 3.6782 & 0.3511 &  5.3521 &  0.4719 &2.4531    & 0.2934 & 3.3905 & 0.3591 &  2.1412& 0.2919& 2.8276 & 0.3199 \\
CasCN & 2.8390 & 0.2914 &4.2041 & 0.3397 & 2.1431  & 0.2852  & 2.8417 & 0.3038 & 1.7607 & 0.3076 & 2.6881 & 0.3593\\
CasFlow & 2.6368 & 0.2630 & 4.3256 & 0.3432 & 1.8548 & 0.2458 & 2.1132 & 0.2372 & 0.9726 & 0.2201 & 1.7537 & 0.2837 \\ %
CTCP & 2.3624 & 0.2799 & 3.8968  & 0.3593  & 1.9860  & 0.2703 & 2.3111 & 0.2830 & 1.2864 &0.2556 &  1.9875 & 0.2760 \\
CasDO & 2.2258 & 0.2673 & 3.6911  & 0.3187  & 1.7286  & 0.2374 & \textbf{2.0549} & \textbf{0.2317} & 1.0286 &0.2219 &  1.6935 & 0.2338 \\ 
\midrule
\textbf{NODEPT} & \textbf{1.7629} & \textbf{0.2635}  & \textbf{3.2276} &\textbf{0.2976} & \textbf{1.6573} & \textbf{0.2251 }& 2.2842  & 0.2576  & \textbf{0.9274} & \textbf{0.2157} & \textbf{1.6792}   & \textbf{0.2183}  \\

\bottomrule
\end{tabular}
}
\label{Model Performance}
\end{table}

\subsection{Performance Comparison (RQ1)}

\subsubsection{Single-point prediction}
To comprehensively verify the effectiveness of NODEPT in single-point prediction task, we compare it with baselines on three datasets. Additionally, we assess the model's performance at two distinct time points for each dataset to gauge both short-term and long-term predictive capabilities separately. The results are shown in Table \ref{Model Performance}, from which the following conclusions can be drawn.

Firstly, NODEPT consistently outperforms other models on the Twitter and APS datasets, validating the effectiveness of our proposed method. Furthermore, NODEPT shows a significant improvement over the state-of-the-art baseline, CasDO, due to its superior ability to capture the dynamics of the underlying diffusion system. On the Weibo dataset, however, NODEPT's advantage is less pronounced compared to CasDO at the 15-hour mark. This discrepancy may be attributed to the relatively short observation period on Weibo, where user interests and the diffusion system undergo less significant evolution.

Secondly, CTCP and CasFlow achieved better performance compared to DeepHawkes and CasCN, indicating that graph-based methods can effectively capture the structural features of cascades and deliver superior results. Meanwhile, CadDO and CasODE achieve relatively better performance by considering the temporal irregularity of cascade events.



\subsubsection{Overall trajectory prediction}

\begin{table}[tbp]
\centering
\caption{Trajectory prediction performance on all datasets}
\begin{tabular}{lcccccc}
\toprule
\multirow{2}{*}{Model} & \multicolumn{2}{c}{Twitter} & \multicolumn{2}{c}{Weibo} & \multicolumn{2}{c}{APS} \\
\cmidrule(lr){2-3} \cmidrule(lr){4-5} \cmidrule(lr){6-7}
& MSLE & MAPE & MSLE & MAPE & MSLE & MAPE \\
\midrule
SEISMIC & 4.4345 & 0.3818 & 4.8685 & 0.4053  & 3.9436 & 0.3538 \\  
CASPER & 3.9408 & 0.3549 &  4.0468  & 0.3505  &3.4609&0.3376 \\ 
\midrule
\textbf{NODEPT} &\textbf{2.3024} & \textbf{0.2645} & \textbf{1.9147} & \textbf{0.2575}&\textbf{1.1493}& \textbf{0.2291 }\\
\bottomrule
\end{tabular}
\label{Trajectory prediction performance}
\end{table}

Previous trajectory prediction models were evaluated using single-point prediction tasks \cite{zhangAnytimeInformationCascade2022}. To assess performance, we initially computed the model's MSLE and MAPE at all integer discrete points within the prediction interval and averaged these values.  The results are presented in Table \ref{Trajectory prediction performance}, leading to the following conclusions. 

NODEPT consistently outperforms both SEISMIC and CASPER on MSLE and MAPE metrics, highlighting its superior ability in overall trajectory prediction.  CASPER, a state-of-the-art baseline for traditional popularity trajectory prediction, assumes that the information diffusion process follows marked Hawkes point processes,
but it fails to fully capture the underlying dynamics driving popularity growth. In contrast, NODEPT leverages neural ODEs to automatically learn complex patterns from data, resulting in excellent predictive performance.



\subsection{Ablation Study (RQ2)}
\begin{table}[htbp]
\centering
\caption{Ablation study results on all datasets}
\begin{tabular}{lcccccc}
\toprule
\multirow{2}{*}{Model} & \multicolumn{2}{c}{Twitter} & \multicolumn{2}{c}{Weibo} & \multicolumn{2}{c}{APS} \\
\cmidrule(lr){2-3} \cmidrule(lr){4-5} \cmidrule(lr){6-7}
& MSLE & MAPE & MSLE & MAPE & MSLE & MAPE \\
\midrule
w/o DL & 4.2475 & 0.3142 & 4.2367 & 0.3136  & 1.2457  & 0.2657 \\  %
w/o TL & 2.9673 &  0.2821 &  2.3115  &  0.3019  &1.3891 &0.3233 \\ 
w/o IR & 3.2887 & 0.2989 & 2.7946 & 0.2858 & 2.1846 & 0.2566\\ 
w/o IL & 2.4741 & 0.2771 & 2.0982 & 0.2736 & 1.1287 & 2.2386 \\
\midrule
NODEPT &2.3024 & 0.2645 & 1.9147 &  0.2575&1.0657& 0.2291 \\
\bottomrule
\end{tabular}
\label{Ablation study}
\end{table}
To investigate the effectiveness of the submodule of NODEPT, we compare it with the following variations. (1) \textbf{w/o DL} removes the dynamic representation learning module. (2) \textbf{w/o TL} removes the temporal representation learning module. (3) \textbf{w/o IR} removes the interaction representation in the ODE generative model. (4) \textbf{w/o IL} removes the increment loss in the Equation \ref{loss}.

To thoroughly evaluate the effectiveness of each module, we computed the evaluation metrics for overall trajectory prediction. The results are presented in Table \ref{Ablation study}, with the observations detailed as follows.
(1) NODEPT outperforms the w/o DL which indicates the dynamic representation learning module can leverage co-evolution structural characteristics for better prediction. (2) The performance degradation of w/o TL shows that the temporal representation learning module can capture the temporal patterns from the diffusion sequence. (3) The w/o IR variant underperforms NODEPT, suggesting that the interaction representation can explicitly capture intricate interactions between cascades through attention mechanism, which is important for tracking the evolution of the underly diffusion system. (4) NODEPT outperforms the w/o IL, indicating that the increment loss can effectively capture the growth pattern of the popularity trajectory.

\subsection{Influence of Hyperparameters (RQ3)}

\begin{figure}
    \centering
    \includegraphics[width=\linewidth]{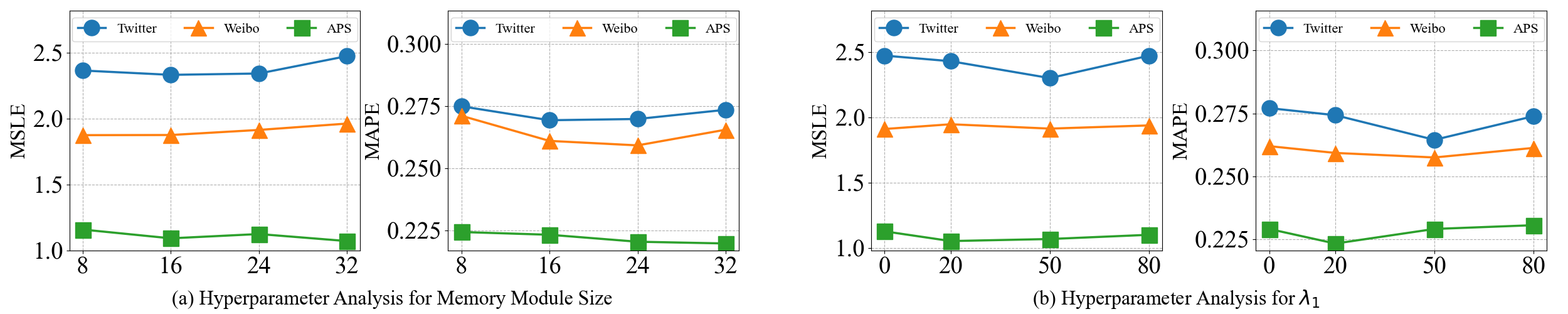}
    \caption{Effect of $\lambda_1$ and memory module size on all datasets.}
    \label{fig:effect of Hyperparameters}
\end{figure}
To analyze the effect of key hyperparameters, we compare the performance of NODEPT under different settings of memory module size and $\lambda_1$ in Equation \ref{loss}. The results are shown in Figure \ref{fig:effect of Hyperparameters}.

\textbf{The Influence of Memory Module Size.} Figure  \ref{fig:effect of Hyperparameters} (a) shows that performance initially improves with an increase in memory module size but starts to decline once the module size becomes too large. This observation highlights the importance of selecting an optimal memory module size that effectively captures relevant interactions while avoiding overfitting or introducing noise into the model predictions.

\textbf{The Influence of $\lambda_1$.} The parameter $\lambda_1$ controls the trade-off between the increment loss and the ELBO loss. Figure \ref{fig:effect of Hyperparameters}  illustrates that the increment loss significantly enhances the model's trajectory prediction performance.
Meanwhile, selecting an appropriate $\lambda_1$ is crucial, as it enables the model to effectively capture both the popularity value and its growth trend.

\subsection{Visualization and Analysis of Popularity Trajectories (RQ4)}

\begin{figure}
    \centering
    \includegraphics[width=\linewidth]{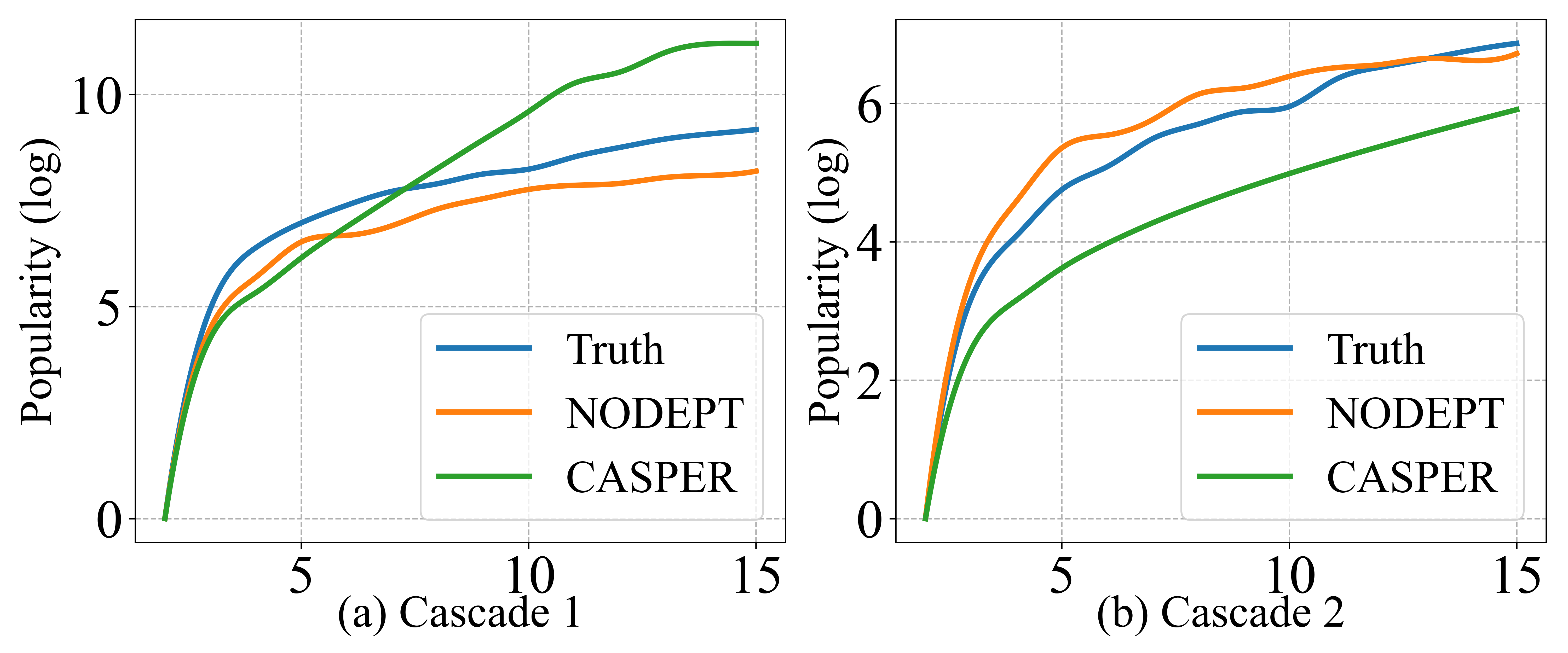}
    \caption{Trajectories of information cascade popularity.}
    \label{fig: Trajectory Visualization}
\end{figure}
To demonstrate the effectiveness of NODEPT for popularity trajectory prediction and compare it with CASPER, we visualized the true and predicted popularity trajectories of both models for two cascades in the Twitter dataset, as shown in Figure \ref{fig: Trajectory Visualization}. Both models perform well in short-term prediction. However, in long-term prediction, NODEPT surpasses CASPER, further validating its ability to effectively model the evolution dynamics of the diffusion system and achieve superior prediction performance.
Additionally, Figure \ref{fig: Trajectory Visualization} shows NODEPT captures the rich dynamic properties of information cascade popularity, including change rates and growth patterns. This capability is essential for understanding the momentum of information diffusion and planning timely interventions.

\section{Conclusion}
In this paper, we propose a novel approach to model the continuous dynamics of diffusion system using neural  ODEs for continuous popularity trajectory prediction. 
We first adopt an encoder which consists of two representation learning modules that capture co-evolution structural characteristics and temporal patterns of cascades from different perspectives for initializing latent state representations of cascades.  More importantly,  we propose an ODE generative module to track the evolution of the underlying diffusion system which effectively models both the self-evolution of individual cascades as well as their intricate interactions with other cascades.  This method addresses the limitations of traditional approaches that rely on fixed diffusion mechanism assumptions, providing more accurate predictions.
 Extensive experiments on three real-world datasets demonstrate the effectiveness and rationality of our proposed  NODEPT approach.

%
%
%
\bibliographystyle{splncs04}
\bibliography{reference}
%




\end{document}